# The Super-Earth Opportunity – Search for Habitable Exoplanets in the 2020s

**Thematic Areas:** ☒ Planetary Systems  ☒ Star and Planet Formation
☐ Formation and Evolution of Compact Objects  ☐ Cosmology and Fundamental Physics
☐ Stars and Stellar Evolution  ☐ Resolved Stellar Populations and their Environments
☐ Galaxy Evolution  ☐ Multi-Messenger Astronomy and Astrophysics

**Principal Author:**
Name: Renyu Hu
Institution: Jet Propulsion Laboratory, California Institute of Technology
Email: renyu.hu@jpl.nasa.gov
Phone: 818-281-9459

**Co-authors:**
A. James Friedson, Christophe Sotin, Mario Damiano, Mark Swain, Michael Mischna, Neal Turner, Pin Chen, Richard Kidd, Robert West, Robert Zellem, Tiffany Kataria, Yasuhiro Hasagawa (Jet Propulsion Laboratory, California Institute of Technology)
Charles A. Beichman (NASA Exoplanet Science Institute, Jet Propulsion Laboratory, California Institute of Technology)
Andrew Howard, Heather Knutson, Yuk Yung (California Institute of Technology)
Chris Reinhard (Georgia Institute of Technology)
Dave Brain (University of Colorado, Boulder)
Edwin Kite, Leslie Rogers (The University of Chicago)
Hilke Schlichting (University of California, Los Angeles)
Noah Planavsky (Yale University)
Rebekah Dawson (Penn State University)
Robert Johnson (University of Virginia)
Sara Seager (Massachusetts Institute of Technology)
Wladimir Lyra (California State University, Northridge)

  

## 1. The Super-Earth Opportunity

The recent discovery of a staggering diversity of planets beyond the Solar System has brought with it a greatly expanded search space for habitable worlds. The *Kepler* exoplanet survey has revealed that most planets in our interstellar neighborhood are larger than Earth and smaller than Neptune. Collectively termed *super-Earths* and *mini-Neptunes*, some of these planets may have the conditions to support liquid water oceans, and thus Earth-like biology, despite differing in many ways from our own planet. In addition to their quantitative abundance, super-Earths are relatively large and are thus more easily detected than true Earth twins. As a result, super-Earths represent a uniquely powerful opportunity to discover and explore a panoply of fascinating and potentially habitable planets in 2020 – 2030 and beyond.

Finding habitable super-Earths is the first step towards finding true Earth analogs. We use the term "habitable" in a conservative sense – a habitable planet is one that can maintain surface liquid water. It is our view that the simplicity of this definition provides useful and conservative guidance in the astronomical search for habitable planets because we rely exclusively on remote sensing to determine exoplanets' atmospheric and surficial compositions. Our use of the term excludes planets with massive gas envelopes (i.e., mini-Neptunes) that would make any planetary surface too hot for the complex molecules required by life.

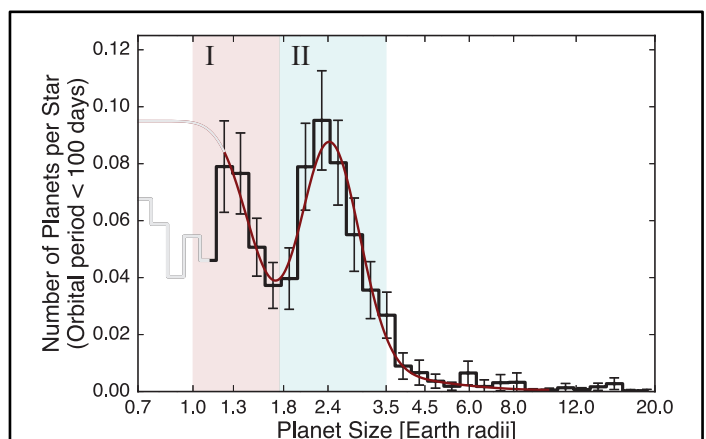

Super-Earths encompass planets with varied formational and evolutionary histories, and a spectrum of diverse compositions. Hundreds of exoplanets now have both mass and radius measurements, which allow basic inferences about their composition. We now know that planets smaller than ~1.7 Earth radii ($R_\oplus$) (i.e., Population I; light red in Fig. 1) likely have predominantly rocky compositions, while larger planets (i.e., Population II; light cyan in Fig. 1) have increasingly greater fractions of gas ($H_2$/He) and/or ice ($H_2O$) (Weiss & Marcy 2014; Rogers 2015). The two populations of planets may be produced by atmospheric loss (Owen & Wu 2017; Jin & Mordasini 2018; Ginzburg et al. 2018).

**Fig. 1.** Occurrence of exoplanets with orbital periods shorter than 100 days derived from the *Kepler* transiting exoplanet survey. The peaks at ~1.3 and ~2.4 $R_\oplus$ suggest two canonical planet types of small planets: the Population I (light red) and the Population II planets (light cyan). The names are not to be confused with those used to describe stellar populations. The two populations of planets have similar intrinsic occurrence frequencies. From Fulton et al. (2017).

Since any planet containing an atmosphere and a liquid water ocean is considered habitable, how can we narrow our search for habitable super-Earths? Planets rocky in their bulk compositions are indeed common (i.e., Population I) and they are obvious targets in the search for habitability. In addition to the bulk composition, the impact of a few wt. % of water upon the planet's upper layer is crucial, despite not adding appreciably to the planet's radius: for example, with a water content less than ~0.2 wt. %, a super-Earth can have an ocean with some land exposed, and such planets can thus be termed "rocky planets." The canonical concepts of the "habitable zone" and "Earth twins" apply to this type of planets. However, only slightly greater water content would lead to full ocean coverage (Cowan & Abbott 2014). If the water content is greater than ~2 wt. %, the bottom of the ocean would have a pressure greater than 2 GPa, and typically high-pressure (HP) ice layers between the crust and the ocean would form (Sotin et al. 2007). There are thus three broad types of habitable super-Earths: (1) rocky planets, (2) shallow ocean planets, and (3) deep ocean planets. Another possible type is a super-Earth with a hydrogen-rich atmosphere, generated by outgassing from the planetary interior (Elkins-Tanton & Seager 2008; Schaefer & Fegley 2010) or via capture of nebular hydrogen during planet formation (Rafikov 2006; Lee et al. 2014; Ginzburg et al. 2016). We refer to this last type as H-rich super-Earths. As a fiducial example, a 10-$M_\oplus$ planet of Earth-like composition would have a radius of 1.75 $R_\oplus$ (e.g., Seager et al. 2007), and a 100-bar $H_2$ atmosphere with a surface temperature of ~300 K would add another 0.25 $R_\oplus$ to the radius (Adams et al. 2008). One may therefore consider any planets that have a radius of less than 2 $R_\oplus$ to be promising targets for habitability if they receive an appropriate amount of irradiation. Furthermore, nature produces "odd balls": we know of rocky planets as large as 2.2 $R_\oplus$ (and 16 $M_\oplus$, Espinoza et al. 2016), and some planets in the Population II planets may well be large ocean planets that do not require massive $H_2$/He envelopes to explain their radius. Given this, the search for habitable super-Earths should encompass the entirety of Population I planets and at least a subset of Population II planets.

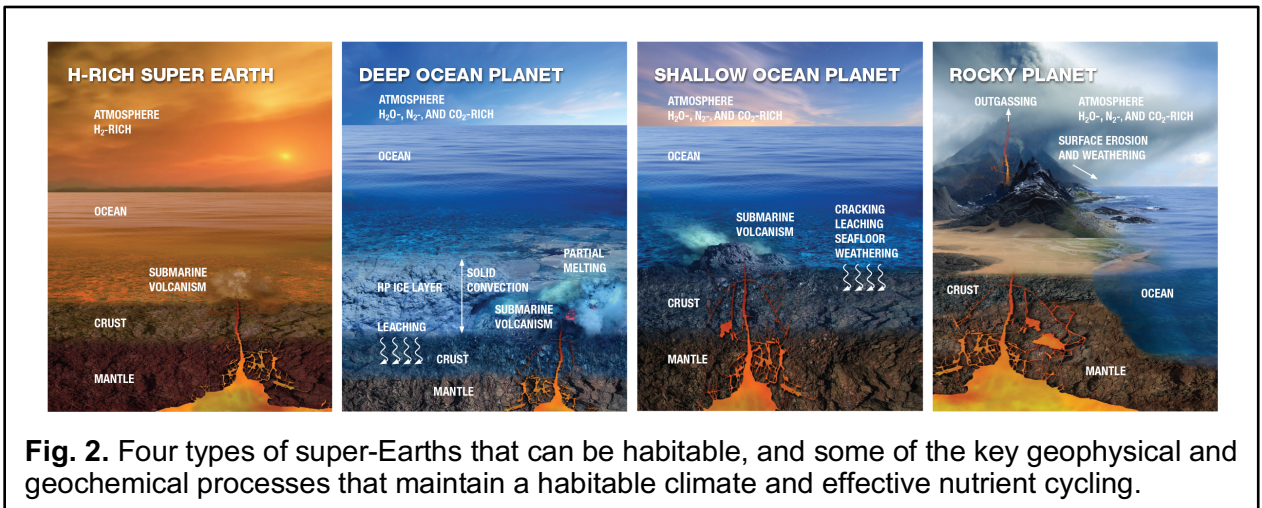

**Fig. 2.** Four types of super-Earths that can be habitable, and some of the key geophysical and geochemical processes that maintain a habitable climate and effective nutrient cycling.

Fig. 2 summarizes the four types of potentially habitable super-Earths. These super-Earths can have a very broad spectrum of atmospheric compositions that would be consistent with surface liquid water, and thus habitability. As such, an Earth-like environment may represent only a narrow fraction of habitable super-Earths. Thus, super-Earths allow us to study habitable exoplanets very different from Earth, and the astronomy and astrobiology communities are ready to seize this "super-Earth opportunity."

## 2. Specific Advantages to Search for Habitability on Super-Earths

Detecting habitable super-Earths will require the characterization of their atmospheres. The mass and the radius of a super-Earth cannot pinpoint its type, because water and gas envelopes can be insignificant contributors to a planet's radius but key contributors to habitability. Super-

3  

Earths are expected to have diverse atmospheric compositions ranging from hydrogen-rich to hydrogen-poor (e.g., Hu & Seager 2014; Morley et al. 2017). To detect atmospheric features from temperate exoplanets, super-Earths have substantial advantages over Earth-sized planets.

First of all, relative to the search for Earth twins, a focus on super-Earths will greatly increase the number of candidate planets where we can search for habitability. As one of the principal goals of the Transiting Exoplanet Survey Satellite (TESS) mission, it will discover a multitude of temperate Earths and super-Earths in transiting orbits of bright stars in the solar neighborhood, so that detailed characterizations of the planets and their atmospheres can be performed. Compared to Earth-sized planets, TESS will yield approximately 3 times more Population I planets and 6 times more Population II planets that receive less than 1.8 times Earth's irradiation, potentially cold enough for water to condense in their atmospheres (Fig. 3).

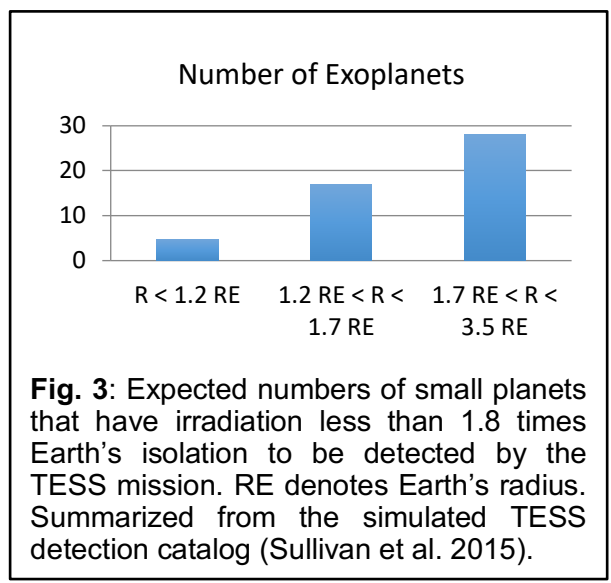

**Fig. 3**: Expected numbers of small planets that have irradiation less than 1.8 times Earth's isolation to be detected by the TESS mission. RE denotes Earth's radius. Summarized from the simulated TESS detection catalog (Sullivan et al. 2015).

Second, H-rich super-Earths orbiting nearby M dwarf stars will provide the eminent opportunity to discover a potentially habitable exoplanet in the first half of the 2020s. Should these planets be discovered by TESS and other exoplanet surveys, the James Webb Space Telescope (JWST) will likely be able to measure their atmospheric transmission spectra from visible to the mid-infrared, which allows detection of $H_2O$, $CH_4$, $NH_3$, and other gases produced by atmospheric photochemical reactions (e.g., Schwieterman et al. 2016). This special type of planets is particularly suitable for study via transmission spectroscopy, because they have atmospheres extended in the vertical scale. For example, as an optimistic estimate, a 1-$R_\oplus$ planet of Earth-like gravity and temperature transiting a 0.15-$R_\odot$ star would have a transmission feature size of 450 ppm for an $H_2$/He-dominated atmosphere, and 43 ppm for a $N_2$-dominated atmosphere. While a 2-$R_\oplus$ planet would have features of 900 ppm and 87 ppm. The spectroscopic features of a temperate $H_2$-dominated atmosphere would thus be well detectable by a single transit. On the contrary, planets with heavier atmospheres will require multiple transits to enhance the S/N to achieve 3σ detection of spectral features (Belu et al. 2011; Schwieterman et al. 2016).

The endeavor to enhance the S/N by stacking multiple transits will critically depend on the following two factors: detector systematics and stellar activities. JWST's near-infrared (NIR) instruments will use the same type of detector as HST's WFC3, and some have suggested that a systematics noise floor of ~20 ppm would be inevitable (Beichman et al. 2014; Greene et al. 2016). This level of noise would prevent JWST to detect non-$H_2$-dominated and temperate atmospheres. On top of this, the host star's spots and faculae cause the wavelength-dependent and changing-with-time effects on the transmission spectrum itself, and this effect is particularly severe for M dwarf stars (Rackham et al. 2018; Zhang et al. 2018). The spectral features of the stellar activities are well above the magnitude of the features from a non-$H_2$-dominated atmosphere, meaning that detecting the atmospheric features would require careful separation of this effect. So far, no methods have been proven to be successful. The detail of this subject is covered in a separate white paper led by B. Rackham.

To characterize habitable super-Earths beyond the H-rich ones, direct imaging will likely be required. For reference, imaging an Earth-like planet around a Sun-like star requires detecting a source $10^{-10}$ fainter than the star, while the requirement for imaging a 2 $R_\oplus$ planet is four times less severe. Therefore, super-Earths can be considerably more favorable targets for direct imaging than Earth-sized planets, for both the random noise (photon counting) and the astrophysical background (exozodiacal dust scattering). Regardless of the specifics of telescopes or instruments, the fundamental limit of the S/N for direct imaging is

$$S/N = \frac{N_P}{\sqrt{N_p + 2(N_S C + N_E + N_Z)}},$$

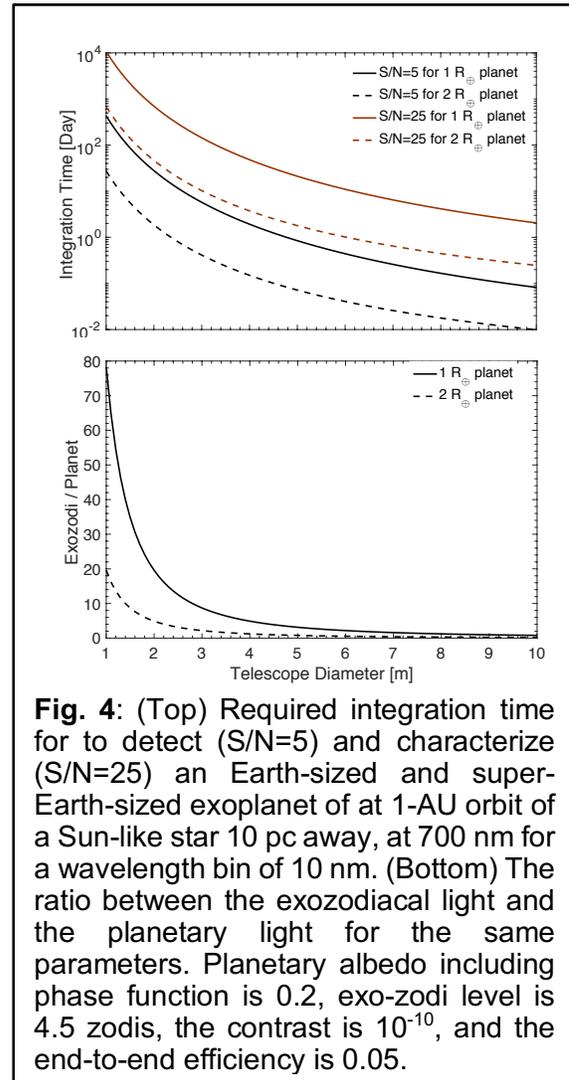

where $N_P$, $N_S$, $N_E$, and $N_Z$ are number of photons of the planet, the star, the exozodiacal dust, and the local zodiacal dust, and $C$ is the contrast. The factor 2 exists because background subtraction is generally needed. This equation omits the terms corresponding to the detector systematics. Fig. 4 shows a reasonable scenario of directly imaging and characterizing planets of a star 10 parsec away. In order to detect the spectral feature of the $O_2$ A band, for a reasonable integration time of ~20 days, an Earth-sized planet requires a 5.2-m telescope, while a 2-$R_\oplus$ planet only requires a 2.6-m telescope. In addition, if we require that the main astrophysical background, $N_E$, would be no more than a factor of 10 greater than the planetary signal $N_P$ for detection, an Earth-sized planet requires a 3.0-m telescope, while 2-$R_\oplus$ planet only requires a 1.4-m telescope. The simple estimates provided here point out a basic fact that directly imaging super-Earths versus directly imaging Earths will require very different, and perhaps generationally different, space telescopes.

**Fig. 4**: (Top) Required integration time for to detect (S/N=5) and characterize (S/N=25) an Earth-sized and super-Earth-sized exoplanet of at 1-AU orbit of a Sun-like star 10 pc away, at 700 nm for a wavelength bin of 10 nm. (Bottom) The ratio between the exozodiacal light and the planetary light for the same parameters. Planetary albedo including phase function is 0.2, exo-zodi level is 4.5 zodis, the contrast is $10^{-10}$, and the end-to-end efficiency is 0.05.

Lastly, another emerging technique is to combine high-contrast imaging (HCI) and high-dispersion spectroscopy (HDS) from ground telescopes to detect atmospheres of exoplanets (e.g., Snellen et al. 2015; Wang et al. 2017). Current HDS instruments on large telescopes can detect CO, $H_2O$, helium, and potentially other molecules from warm and hot giant planets (e.g., Snellen et al. 2010; Lockwood et al. 2014; Allart et al. 2018). Combining HCI with HDS on large and future extremely large telescopes (e.g., E-ELT, 40-m primary mirror) may enable detection of the signal due to biogenic molecules (e.g. $O_2$ bands at 0.76 μm and 1.26 μm) from Earth-sized planets of nearby M stars like Proxima Centauri b (Lovis et al. 2017; Wang et al. 2017).

## 3. Models to Guide the Super-Earth Habitability Search

To guide the search and characterization of super-Earths, which have no analogs in the Solar System, will require a first-principles approach. Previous studies of exoplanet habitability have focused on Earth-sized planets, drawing extensive analogy to Earth (e.g., Kasting et al. 1993, 2014; Kaltenegger et al. 2010; Robinson et al. 2011; Kopparapu et al. 2013; Misra et al. 2014; Schwieterman et al. 2016). However, we must treat super-Earths as unique astrophysical objects since they can have very different atmospheric and interior compositions and formational and evolutionary histories than Earth. Models in the next decade will strive to determine the ranges of habitable environments that are geologically reasonable, which could differ from the habitable environments observed on Earth, by integrating the physical and chemical processes from formation, to evolution, to large-scale atmosphere-ocean-rock chemistry. Key scientific questions that would be addressed by theoretical and modeling studies include

- What are the conditions that form volatile-rich super-Earths, with ices such as $H_2O$ and/or gases such as $H_2$ and He?
- How do super-Earth retain and distribute volatiles to produce habitable environments? For instance, under what conditions a rocky planet can evolve to the state of an $N_2$-$CO_2$-$H_2O$ atmosphere required by the volcanism-weathering cycle?
- What types of atmospheres and circulation patterns that maintain habitable environments on super-Earths?
- What are the spectral and temporal features of each type of habitable super-Earths?

## 4. Recommendations

In this paper we argue that the science return of detecting temperate exoplanets as large as 2 $R_\oplus$ should be considered as one of the primary evaluation criteria to assess the options of exoplanet missions for the next decade. Given the larger size of super-Earths, their potential atmospheric diversity, and strong implications on the size of direct imaging space telescopes, this evaluation criteria should be applied separately and independently from the assessment of a mission or instrument's capability to find a true Earth analog.

We additionally argue that NASA should continue to support theoretical, modeling, and computational research of the characteristics and evolution of habitable super-Earths, drawing from the science community's multi-disciplinary expertise. As we will have a significant sample of atmospheric compositions for super-Earths well before Earth-like planets, and super-Earths are not simply scaled-up Earths but are unique astrophysical objects, the success of any exoplanet characterization projects will essentially depend on our understanding of how super-Earths develop and sustain their habitability.

## Acknowledgement

The research was carried out at the Jet Propulsion Laboratory, California Institute of Technology, under a contract with the National Aeronautics and Space Administration.